\documentclass[12pt,aps,prb,preprint]{revtex4}

\usepackage{amsmath}
\usepackage{amsfonts}
\usepackage{amssymb}
\input{Qcircuit}

\begin{document}
\title{A Discussion on the Teleportation Protocol for States of $N$
Qubits.}

\author{Alejandro D\'{\i}az-Caro}
    \affiliation{Departamento de Ciencias de la Computaci\'on. Facultad
de Ciencias Exactas, Ingenier\'{\i}a y Agrimensura, Universidad
Nacional de Rosario, Pellegrini 250, 2000 Rosario, Argentina}
    \email{diazcaro@fceia.unr.edu.ar}
\author{Manuel Gadella}
    \affiliation{Departamento de F\'{\i}sica Te\'orica. Facultad de
Ciencias, Universidad de Valladolid, c. Real de Burgos, s.n.,
47011 Valladolid, Spain}.
    \email{gadella@fta.uva.es}
\date{\today}

\begin{abstract}
In this paper, we want to present a simple and comprehensive
method to implement teleportation of a system of $N$ qubits and
its discussion on the corresponding quantum circuit. The paper can
be read for nonspecialists in quantum information.
\end{abstract}

\maketitle

\section{Motivation}\label{sec:Motivation}

Quantum Information and quantum computation are subjects that are
recently received an enormous interest in the scientific
community. Textbooks of a high pedagogical value have been written
on these subjects. Among them, we can quote the monographs from
Nielsen and Chuang \cite{NC} and Preskill \cite{P}. Nevertheless,
some important procedures, which in addition have not been treated
in the monographs, need a reformulation and a presentation that
make them accessible to the physics teacher.

We have chosen the teleportation of $N$ qubits because i.) this is
a relevant subject. In fact, once we have an algorithm that can
teleport a quantum state, we immediate ask for a natural and
simple generalization that can enable us to transmit a large
amount of information at long distance. ii.) protocols to teleport
$N$ qubit states have already been published \cite{R}. We want to
introduce here another $N$ qubit teleportation protocol based in
the previous ones \cite{R}, but written in a simpler new fashion
that pretends to be more useful and more clear to the average
physicists interested in these subjects.

The possibility of teleportation of a qubit state has been
suggested in 1993 by Bennet et al \cite{B}. Later, in 1998,
Brassard developed a quantum circuit in order to implement one
qubit teleportation \cite{BR}. In the description of the
teleportation of an $N$ qubit state, we shall also introduce a
circuit that will do the job. The advantage of the circuit
notation is that it makes easier the comprehension of the process
through a visualization of it.

The protocol for the teleportation of one qubit state is very well
known and has been discussed in textbooks as for instance in
monograph from Nielsen and Chuang \cite{NC}. The best known
teleportation protocol can be summarized in few words as follows:
Let us start with the qubit that we want to teleport, represented
by the state $|\psi\rangle=\alpha|0\rangle+\beta|1\rangle$ where
$|0\rangle$ denotes \cite{7} ``spin up'' and $|1\rangle$ ``spin
down'', and let us consider the auxiliar two qubit state
represented by the Bell state:

\begin{equation}\label{1}
|\beta_{00}\rangle= \frac{|00\rangle+|11\rangle}{\sqrt 2}\,.
\end{equation}

Let us consider the following three qubit state:

\begin{equation}\label{2}
|\psi\rangle|\beta_{00}\rangle= \frac{1}{\sqrt
2}\{\alpha|0\rangle[|00\rangle+|11\rangle]+\beta|1\rangle
[|00\rangle+|11\rangle]\}\,.
\end{equation}
The first qubit correspond to the state that we want to teleport.
This and the second qubit are assumed to be and to remain in the
hands of Alice, the sender. The third qubit is brought by Bob, the
receiver, to his location. Alice can thus manipulate her two
qubits and she does it as follows: First, she makes them to pass
through a CNOT gate. The CNOT gate flips the second qubit if the
first one is in the state $|1\rangle$ and keeps it unchanged if
the former is $|0\rangle$.

Then, Alice makes her first qubit passing through a Hadamard gate.
We recall that a Hadamard gate produces the following changes:

\begin{equation}\label{3}
    |0\rangle\longmapsto \frac{1}{ 2}\{ |0\rangle+|1\rangle\}
    \hskip0.5cm;\hskip0.5cm |1\rangle\longmapsto \frac{1}{\sqrt 2}\{
    |0\rangle-|1\rangle\}\,.
\end{equation}

Thus, the resulting three qubits state can be written as
follows\cite{NC}:

\begin{eqnarray}
 |\psi_I\rangle= \frac 12
 \{|00\rangle(|\alpha|0\rangle+\beta|1\rangle)
 +|01\rangle(\alpha|1\rangle+\beta|0\rangle)\nonumber\\[2ex]
+|10\rangle
 (\alpha|0\rangle-\beta|1\rangle)+|11\rangle
 (\alpha|1\rangle-\beta|0\rangle)\}\,.\label{4}
\end{eqnarray}

Then, Alice produces a measurement on her two qubits. She can have
one out of four results only:
$\{|00\rangle,|01\rangle,|10\rangle,|11\rangle\}$. By means of an
open classical system of communication (telephone, e-mail, etc),
Alice communicates to Bob the result obtained. Accordingly, Bob
produces an operation on his qubit so as to obtain the original
state (Since this operation is performed in an environment
different from the original lab where the first qubit was
produced, we can say that the original state has been teleported).
This operation is:

\begin{center}
\begin{tabular}{|c|c|}
    \hline
    Alice result & Bob's operation on his qubit\\
    \hline
    $|00\rangle$ & Does nothing \\
    $|01\rangle$ & X gate \\
    $|10\rangle$ & Z gate \\
    $|11\rangle$ & ZX gate\\
    \hline
\end{tabular}
\end{center}

We recall that the $X$ and $Z$ gate are the $\sigma_x$ and
$\sigma_z$ Pauli matrices respectively \cite{8}. The $ZX$ gate
means that we first apply the $X$ gate and then the $Z$ gate, this
notation is copied from the usual algebraic manipulation according
to which the first operation lies in the right the second on its
left and so on. All these operations can be written as
$Z^{M_1}X^{M_2}$, where $M_i$, with $i=1,2$, are either $0$ or
$1$.

Thus, if for example the measurement of the first two qubits give
$|00\rangle$, the third qubit must be already in the state we want
teleport as shown in (\ref{4}). Bob does not need to do anything
as he has the wanted state. This operation {\it do nothing} is
written in algebraic form as $Z^0X^0$. If Alice gets $|01\rangle$,
then Bob applies $X\equiv Z^0X^1$ to his qubit, etc.

Note that the gate $X$ is equivalent to the NOT gate that produces
the flips $|0\rangle\longmapsto |1\rangle$ and
$|1\rangle\longmapsto |0\rangle$, while $Z$ just changes the sign
in front of $|1\rangle$ \cite{NC}.

It is noteworthy to recall the obvious fact that the subject of
quantum teleportation are qubit states and not any kind of
particles with or without mass (qubit could be implemented using
photon polarization).

The remainder of this paper is organized as follows: In the brief
Section \ref{sec:UsefulNotations}, we introduce a useful notation
that helps in abbreviating lengthly formulas. It is in the long
section \ref{sec:TeleportationAlgorithm} where we discuss the
teleportation algorithm. The algorithm is obtained by induction on
the number of qubits of the state to be teleported. This induction
procedure is discussed in detail in subsection
\ref{sec:ProofOf12}. Previously, in subsection
\ref{sec:ProofOf11}, we have derived an important intermediate
formula. Finally, on subsection \ref{sec:DescripCircuit}, we
present the quantum circuit for the teleportation with careful
explanation of all its constituents.

The paper is written in a style that can be read for physicists,
quantum chemists and computer scientists without previous
experience in quantum information theory, thus being intended for
a wide audience.

\section{Useful Notations.}\label{sec:UsefulNotations}

As we intend a thoroughly discussion on a rather cumbersome
manipulation as teleportation of $N$ qubits is and, at the same
time, we pretend to give an accesible version of this operation,
it seems natural to search for a notation that can help us in our
goal. To this end, we choose the following:

i.) Let us denote by $k_n$ a chain of $n$ bits, where $k$ is a
natural number and $k_n$ is the chain that represents $k$ in terms
of these $n$ bits.

ii.) One of the tools available is the so called Iverson delta. In
order to define it, we first associate to each property $p$ a
number, as introduced by Iverson \cite{KI}, such that

\begin{equation}\label{5}
    [p]:=\left\{
\begin{array}{ccc}
  1& {\rm if} & p \\[2ex]
  0 & {\rm if} & \sim p \\
\end{array}
    \right.\,.
\end{equation}
This is called the Iverson notation \cite{KI}. After (\ref{5}), we
define the Iverson delta as follows: let $i$ and $k$ two natural
numbers with chain of digits $k_n$ and $i_n$. Then,

\begin{equation}\label{6}
    \ddot \delta_{i,k}:= [\,(i_n \;{\rm AND}\; k_n)\;\; {\rm have\;
    an\;
    odd\; number\; of\; bits\; with\; the\; digit\; 1}\,]\,.
\end{equation}
We are listing below some interesting properties of the Iverson
delta. Their proof  is not essential in our presentation (and
otherwise easy to obtain) and we omit it here:

\medskip
1.- $(-1)^{\ddot \delta_{2i+1,2k+1}}=(-1)^{\ddot\delta_{i,k}+1}$.

\medskip
2.- $(-1)^{\ddot\delta_{2i+1,2k}}=(-1)^{\ddot\delta_{i,k}}$.

\medskip
3.- $(-1)^{\ddot\delta_{2i,2k+1}}=(-1)^{\ddot\delta_{i,k}}$.

\medskip
4.- $(-1)^{\ddot\delta_{2i,2k}}=(-1)^{\ddot\delta_{i,k}}$.

\medskip
iii.) This kind of replacement is very usual:

\begin{equation}\label{7}
    \sum_{a_1\dots a_n=k}^l\equiv \sum_{a_1=k}^l\cdots
    \sum_{a_n=k}^l\,.
\end{equation}

So far the explanation of notation to be used in the sequel. In
the next section, we start with our presentation.

\section{Teleportation Algorithm.}\label{sec:TeleportationAlgorithm}

We want to teleport an arbitrary pure state of $N$ qubits that we
shall denote by $|\psi_N\rangle$. Then, $|\psi_N\rangle$ is a
vector state of the tensor product of $N$ times the two
dimensional Hilbert space ${\mathbb C}^2$, where qubits dwell
\cite{9}.

Once the state $|\psi_N\rangle$ has been prepared, we need a
device to teleport it. In the case of $N=1$, we have seen that $2$
additional or auxiliary (also called ancillary) qubits are needed.
In our case, we can expect that we shall require $2N$ auxiliary
qubits. Then, we have $3N$ qubits that we shall distribute into
three groups.

First of all the $N$ qubits whose state we want to teleport. These
qubits will make the first group and its state denoted as
$|\psi_N\rangle_1$, where we have added the subindex $1$
accordingly. The state $|\psi_N\rangle_1$ can be written in terms
of the $N$ qubit basis $|i_N\rangle$ as

$$
|\psi_N\rangle_1= \sum_{i=0}^{2^N-1}\alpha_i\,|i_N\rangle_1\,.
$$
We assume that $|\psi_N\rangle_1$ is normalized.

The second group will be formed by the first $N$ auxiliary qubits.
Before the beginning of the teleportation procedure all them are
prepared to be in the state $|0\rangle$. Then, the quantum state
for the system of these $N$ qubits is $|00\dots 0\rangle_2$ with
$N$ zeroes. If we use the notation described in the previous
section, a chain of $N$ zeros is described by $0_N$, so that the
state of this second group of qubits is here denoted as
$|0_N\rangle_2$.

By the same arguments, we write the initial state of the third
group of qubits as $|0_N\rangle_3$.

The teleportation protocol for a $N$ qubit state can be looked as
a generalization of the $N=1$ case. With this idea in mind, let us
take the first $N$ auxiliar qubits in the collective state
$|0_N\rangle_2$ and let them pass through respective Hadamard
gates.

Once this operation has been completed, take the auxiliary qubits
of the third group and make the following CNOT operations; the
$(N+1)$-th auxiliary qubit (after Hadamard!) with the $(2N+1)$-th,
the $(N+2)$-th with the $(2N+2)$-th and so on. See Figure 1.

The final result is the complete entanglement of the $2N$
auxiliary qubits. The collective state of the system of all these
qubits resulting after these manipulations is a generalization of
the Bell state (\ref{1}) used in the $N=1$ case. This generalized
Bell state has the following form:

\begin{equation}\label{8}
    \frac{1}{\sqrt{2^N}}\sum_{j=0}^{2^N-1} |j_Nj_N\rangle_{23}\,.
\end{equation}
For instance, for $N=2$, this sum gives:

\begin{equation}\label{9}
    \frac12\{|0000\rangle+|0101\rangle+|1010\rangle
    +|1111\rangle\}\,.
\end{equation}

At this point, Alice (sender) and Bob (receiver) move away from
each other. The generalization of the $N=1$ case suggests that
Alice keeps the $2N$ first qubits, i.e., the $N$ qubit state to be
teleported $|\psi_N\rangle_1$ and the first $N$ auxiliary qubits
in their final state $|\ \rangle_2$.

Next, Alice performs a CNOT operation between the qubit $k$ and
the qubit $k+N$ for all $k=1,2,\dots,N$ (see Figure 1). This
produces the following state of the $3N$ qubit system:

\begin{equation}\label{10}
    \frac{1}{\sqrt{2^N}}\sum_{i=0}^{2^N-1}\alpha_i\,|i_N\rangle_1
    \sum_{j=0}^{2^N-1}|(j\;{\rm XOR}\; i)_Nj_N\rangle_{23}\,.
\end{equation}
We recall that the operation \cite{10} $j$ XOR $i$ means $j+i$
modulus $2$, i.e., $0+0=0$, $0+1=1+0=1$ and $1+1=0$. For example,
if $N=2$ and $j=2$ ($j_2=10$) and $i=3$ ($i_2=11$), we have that
$(j\;{\rm XOR}\; i)_N=01\equiv 1$.

The next operation performed by Alice is applying a Hadamard gate
to each of the first $N$ qubits (which original state we want to
teleport, see Figure 1). In subsection \ref{sec:ProofOf11}, we
show that the final state of the $3N$ qubit system is given by

\begin{equation}\label{11}
    \frac{1}{2^N}\sum_{i,j,k=0}^{2^N-1}|k_N(j\;{\rm
XOR}\;i)_N\rangle_{12}\,(-1)^{\ddot\delta_{i,k}}\alpha_i\,|j_N\rangle_3\,.
\end{equation}

Nevertheless, (\ref{11}) is not the most useful form of the state
of this entanglement of $3N$ qubits. In subsection
\ref{sec:ProofOf12}, we shall show that formula (\ref{11}) is
equal to

\begin{equation}\label{12}
    \frac{1}{2^N}\sum_{a_1\cdots a_N=0}^1 |a_1\cdots
    a_{2N}\rangle_{12}\left(\bigotimes_{k=a_{N+1}}^{a_{2N}}X^k\right)
    \left(\bigotimes_{l=a_1}^{a_N}Z^l\right)|\psi_N\rangle_3\,.
\end{equation}

Then, Alice makes a measurement of her $2N$ qubits. Assume that
the result is $|a_1\cdots a_{2N}\rangle_{12}$. Then, the state of
the $N$ quibits own by Bob is given by

\begin{equation}\label{13}
|\varphi\rangle_3=\left(\bigotimes_{k=a_{N+1}}^{a_{2N}}X^k\right)
    \left(\bigotimes_{l=a_1}^{a_N}Z^l\right)|\psi_N\rangle_3\,.
\end{equation}

In order to obtain the original state $|\psi_N\rangle$, Bob must
multiply $|\varphi\rangle_3$ in (\ref{13}) by the inverse of the
operator $\left(\otimes_{k=a_{N+1}}^{a_{2N}}X^k\right)
    \left(\otimes_{l=a_1}^{a_N}Z^l\right)$.  Hence,
    teleportation of $|\psi_N\rangle$ is completed.

Although formulas (\ref{11}) and (\ref{12})  describe the same
$3N$ qubit entangled state, we see that (\ref{11}) is useless for
teleportation while the usefulness of (\ref{12}) is quite obvious.
However, the derivation of (\ref{12}) from (\ref{11}) is not
immediate and needs some discussion. This is presented in
subsection \ref{sec:ProofOf12}.

We recall that the whole procedure is described by a circuit. This
is presented in Figure 1 and subsection \ref{sec:DescripCircuit}.

\subsection{Proof of (\ref{11}).}\label{sec:ProofOf11}

Our next goal is to show that (\ref{10}) plus the operation of
passing the first $N$ qubits through respective Hadamard gates
gives (\ref{11}). First of all, let us consider the state
$|i_N\rangle$ of a system of $N$ qubits. Each qubit, will pass
through a Hadamard gate, this action is represented as $H^{\otimes
N}$, where $H$ stands for Hadamard gate. Then we have to show that

\begin{equation}\label{14}
H^{\otimes N}|i_N\rangle =
\frac{1}{\sqrt{2^N}}\sum_{k=0}^{2^N-1}\,(-1)^{\ddot\delta_{i,k}}|k_N\rangle\,.
\end{equation}

We shall prove this result by induction on $N$. For $N=2$, we call
$x$ and $y$ to the first and second qubit. We use the properties
of the Iverson delta. Then,

\begin{eqnarray}
  H^{\otimes 2}|xy\rangle =H|x\rangle\,H|y\rangle=\frac{1}{\sqrt 2}
  (|0\rangle+(-1)^x|1\rangle)\,\frac{1}{\sqrt 2}
  (|0\rangle+(-1)^y|1\rangle)\nonumber \\
  =\frac 12 [|00\rangle+(-1)^y|01\rangle+(-1)^x|10\rangle +(-1)^{x+y}
|11\rangle ]\nonumber \\
 =\frac 12 \sum_{k=0}^3 (-1)^{\ddot\delta_{xy,k}} |k_2\rangle\,,
 \label{15}
\end{eqnarray}
which proves (\ref{14}) for $N=2$.

Now, we assume that the result is true for $N=3,4,\dots,n$. Under
this hypothesis, if we prove it for $N=n+1$ it would be shown for
any value of $N$ by induction. We start with the $n+1$ qubit state
$|i_{n+1}\rangle\equiv |j_n\rangle\,|x\rangle$ and make each qubit
pass through a Hadamard gate:

\begin{eqnarray}
 H^{\otimes n+1}|i_{n+1}\rangle= H^{\otimes n+1} |j_n\rangle\,|x\rangle
=
 H^{\otimes n} |j_n\rangle\, H|x\rangle \nonumber \\[2ex]
 =\left[
\frac{1}{\sqrt{2^n}}\sum_{k=0}^{2^n-1}(-1)^{\ddot\delta_{j,k}}|k_n\rangle\right]\,\frac{1}{\sqrt
 2}\,
(|0\rangle+(-1)^x|1\rangle) \nonumber  \\[2ex]
=  \frac{1}{\sqrt{2^{n+1}}}  \sum_{k=0}^{2^n-1}\left(
(-1)^{\ddot\delta_{j,k}}|k_n\rangle |0\rangle+
(-1)^{\ddot\delta_{x,k}}(-1)^x|k_n\rangle|1\rangle\right)
\nonumber\\[2ex]
  = \frac{1}{\sqrt{2^{n+1}}} \sum_{k=0}^{2^{n+1}-1}
(-1)^{\ddot\delta_{j,k}}|k_{n+1}\rangle\,.\label{16}
\end{eqnarray}

Then, let us go back to (\ref{10}), and make pass the first $N$
qubits through respective Hadamard gates. If we call $H^{\otimes
N}$ this operation, the result is

\begin{eqnarray}
  H^{\otimes N}\left(
\frac{1}{\sqrt{2^N}}\sum_{i=0}^{2^N-1}\alpha_i\,|i_N\rangle_1
    \sum_{j=0}^{2^N-1}|(j\;{\rm XOR}\; i)_Nj_N\rangle_{23}\right)
\nonumber
    \\[2ex]
= \frac{1}{\sqrt{2^N}}\sum_{i=0}^{2^N-1}\alpha_i\,
\left[H^{\otimes N} \,|i_N\rangle_1 \sum_{j=0}^{2^N-1}|(j\;{\rm
XOR}\; i)_Nj_N\rangle_{23}\right] \nonumber
\\
  = \frac{1}{\sqrt{2^N}} \sum_{i=0}^{2^N-1}\alpha_i\,\left[\left(
\frac{1}{\sqrt{2^N}} \sum_{k=0}^{2^N-1}
(-1)^{\ddot\delta_{i,k}}|k_N\rangle_1\right) \sum_{j=0}^{2^N-1}
|(j\;{\rm XOR}\; i)_Nj_N\rangle_{23}\right] \nonumber
\\[2ex]
 = \frac{1}{2^N}\sum_{i,j,k=0}^{2^N-1}|k_N(j\;{\rm
XOR}\;i)_N\rangle_{12}\,(-1)^{\ddot\delta_{i,k}}\alpha_i\,|j_N\rangle_3\,.
\label{17}
\end{eqnarray}
Observe that we have written between parenthesis in the third row
in (\ref{17})  the action of $N$ Hadamard gates on the state of
the first $N$ qubits, i.e., $H^{\otimes N}|i_N\rangle_1$. This
ends the proof (\ref{10})$\Longrightarrow$(\ref{11}).

\subsection{Proof of (\ref{12}).}\label{sec:ProofOf12}

In order to show our claim, we shall make use again of an argument
based in the induction principle. Thus, we begin with the proof of
(\ref{12}) for the simplest case of $N=2$, i.e., with the
situation involving two qubits only. In this case, the general
form of a two qubit state is given by

\begin{equation}\label{18}
    |\psi_2\rangle=
\alpha_0|00\rangle+\alpha_1|01\rangle+\alpha_2|10\rangle+\alpha_3|11\rangle\,,
\end{equation}
where $\alpha_i$, $i=0,1,2,3$ are complex numbers such that
$\sum_{i=0}^3 |\alpha_i|^2=1$.

Thus, let us display (\ref{11}) for $N=2$:

\begin{eqnarray}
\frac{1}{4}\sum_{i,j,k=0}^{3}|k_2(j\;{\rm
XOR}\;i)_2\rangle_{12}\,(-1)^{\ddot\delta_{i,k}}\alpha_i\,|j_2\rangle_3\nonumber\\[3ex]
    =\frac{1}{4} [ |0000\rangle_{12}\;
(\alpha_0|00\rangle_3+\alpha_1|01\rangle_3+\alpha_2|10\rangle_3+\alpha_3|11\rangle_3) \nonumber\\
    + |0001\rangle_{12}\;
(\alpha_0|01\rangle_3+\alpha_1|00\rangle_3+\alpha_2|11\rangle_3+\alpha_3|10\rangle_3) \nonumber\\
    + |0010\rangle_{12}\;
(\alpha_0|10\rangle_3+\alpha_1|11\rangle_3+\alpha_2|00\rangle_3+\alpha_3|01\rangle_3) \nonumber\\
    + |0011\rangle_{12}\;
(\alpha_0|11\rangle_3+\alpha_1|10\rangle_3+\alpha_2|01\rangle_3+\alpha_3|00\rangle_3) \nonumber\\
    + |0100\rangle_{12}\;
(\alpha_0|00\rangle_3-\alpha_1|01\rangle_3+\alpha_2|10\rangle_3-\alpha_3|11\rangle_3) \nonumber\\
    + |0101\rangle_{12}\;
(\alpha_0|01\rangle_3-\alpha_1|00\rangle_3+\alpha_2|11\rangle_3-\alpha_3|10\rangle_3) \nonumber\\
    + |0110\rangle_{12}\;
(\alpha_0|10\rangle_3-\alpha_1|11\rangle_3+\alpha_2|00\rangle_3-\alpha_3|01\rangle_3) \nonumber\\
    + |0111\rangle_{12}\;
(\alpha_0|11\rangle_3-\alpha_1|10\rangle_3+\alpha_2|01\rangle_3-\alpha_3|00\rangle_3) \nonumber\\
    + |1000\rangle_{12}\;
(\alpha_0|00\rangle_3+\alpha_1|01\rangle_3-\alpha_2|10\rangle_3-\alpha_3|11\rangle_3) \nonumber\\
    + |1001\rangle_{12}\;
(\alpha_0|01\rangle_3+\alpha_1|00\rangle_3-\alpha_2|11\rangle_3-\alpha_3|10\rangle_3) \nonumber\\
    + |1010\rangle_{12}\;
(\alpha_0|10\rangle_3+\alpha_1|11\rangle_3-\alpha_2|00\rangle_3-\alpha_3|01\rangle_3) \nonumber\\
    + |1011\rangle_{12}\;
(\alpha_0|11\rangle_3+\alpha_1|10\rangle_3-\alpha_2|01\rangle_3-\alpha_3|00\rangle_3) \nonumber\\
    + |1100\rangle_{12}\;
(\alpha_0|00\rangle_3-\alpha_1|01\rangle_3-\alpha_2|10\rangle_3+\alpha_3|11\rangle_3) \nonumber\\
    + |1101\rangle_{12}\;
(\alpha_0|01\rangle_3-\alpha_1|00\rangle_3-\alpha_2|11\rangle_3+\alpha_3|10\rangle_3) \nonumber\\
    + |1110\rangle_{12}\;
(\alpha_0|10\rangle_3-\alpha_1|11\rangle_3-\alpha_2|00\rangle_3+\alpha_3|01\rangle_3) \nonumber\\
    + |1111\rangle_{12}\;
(\alpha_0|11\rangle_3-\alpha_1|10\rangle_3-\alpha_2|01\rangle_3+\alpha_3|00\rangle_3)
    ] \label{19}
\end{eqnarray}

After an easy but rather cumbersome term by term analysis of the
previous sum, we show that (\ref{19}) is equal to

\begin{equation}\label{20}
    \frac14\sum_{a,b,c,d=0}^1|abcd\rangle_{12}(X^c\otimes
    X^d)(Z^a\otimes Z^b)\,|\psi_2\rangle_3\,,
\end{equation}
where, as in the case $N=1$, $X\equiv \sigma_x$ and $Z\equiv
\sigma_z$, the Pauli matrices \cite{11}. Note that if $M$ is any
Pauli matrix, one has that $M^0=I$, the identity matrix. Of
course, $M^1=M$.

Once we have proven our result for $N=2$, the induction procedure
assumes that the same is true for $N=3,\dots, n$. Then, if we
prove that the result is true for $N=n+1$, it is proven for any
natural number $N$. Then, let us take the $n+1$ qubit state given
by

\begin{equation}\label{a}
    |\psi_{n+1}\rangle =\sum_{i=0}^{2^{n+1}-1}\alpha_i|i_{n+1}\rangle
\end{equation}

To this end, we need to write formula (\ref{11}) for $N=n+1$ and
to span it in a sum as follows:

\begin{eqnarray}
 \frac{1}{2^{n+1}}\sum_{i,j,k=0}^{2^{n+1}-1} |k_{n+1}\;(j\;\;{\rm
 XOR}\;\;
  i)_{n+1}\rangle_{12}  (-1)^{\ddot
\delta_{i,k}}\,\alpha_i\,|j_{n+1}\rangle_3\nonumber
  \\[3ex] =\frac 12\left[\frac{1}{2^n} \sum_{i,j,k=0}^{2^{n}-1}
|k_{n}\,0\;(j\;\;{\rm
 XOR}\;\;
  i)_{n}\,0\rangle_{12}  (-1)^{\ddot
\delta_{2i,2k}}\,\alpha_{2i}\,|j_{n}\,0\rangle_3
  \right.\label{21}\end{eqnarray}
  \begin{eqnarray} +
\frac{1}{2^n} \sum_{i,j,k=0}^{2^{n}-1} |k_{n}\,0\;(j\;\;{\rm
 XOR}\;\;
  i)_{n}\,1\rangle_{12}  (-1)^{\ddot
\delta_{2i+1,2k}}\,\alpha_{2i+1}\,|j_{n}\,0\rangle_3
  \label{22}\\
  +
\frac{1}{2^n} \sum_{i,j,k=0}^{2^{n}-1} |k_{n}\,1\;(j\;\;{\rm
 XOR}\;\;
  i)_{n}\,0\rangle_{12}  (-1)^{\ddot
\delta_{2i,2k+1}}\,\alpha_{2i}\,|j_{n}\,0\rangle_3
  \label{23}\end{eqnarray}
  \begin{eqnarray}
  +
\frac{1}{2^n} \sum_{i,j,k=0}^{2^{n}-1} |k_{n}\,1\;(j\;\;{\rm
 XOR}\;\;
  i)_{n}\,1\rangle_{12}  (-1)^{\ddot
\delta_{2i+1,2k+1}}\,\alpha_{2i+1}\,|j_{n}\,0\rangle_3
\label{24}\\
  +
\frac{1}{2^n} \sum_{i,j,k=0}^{2^{n}-1} |k_{n}\,0\;(j\;\;{\rm
 XOR}\;\;
  i)_{n}\,1\rangle_{12}  (-1)^{\ddot
\delta_{2i,2k}}\,\alpha_{2i}\,|j_{n}\,1\rangle_3
 \label{25}\end{eqnarray}
  \begin{eqnarray}
  +
\frac{1}{2^n} \sum_{i,j,k=0}^{2^{n}-1} |k_{n}\,0\;(j\;\;{\rm
 XOR}\;\;
  i)_{n}\,0\rangle_{12}  (-1)^{\ddot
\delta_{2i+1,2k}}\,\alpha_{2i+1}\,|j_{n}\,1\rangle_3
  \label{26}\\
  +
\frac{1}{2^n} \sum_{i,j,k=0}^{2^{n}-1} |k_{n}\,1\;(j\;\;{\rm
 XOR}\;\;
  i)_{n}\,1\rangle_{12}  (-1)^{\ddot
\delta_{2i,2k+1}}\,\alpha_{2i}\,|j_{n}\,1\rangle_3
 \label{27}\\
  \left.+
\frac{1}{2^n} \sum_{i,j,k=0}^{2^{n}-1} |k_{n}\,1\;(j\;\;{\rm
 XOR}\;\;
  i)_{n}\,0\rangle_{12}  (-1)^{\ddot
  \delta_{2i+1,2k+1}}\,\alpha_{2i+1}\,|j_{n}\,1\rangle_3\right]
  \label{28}
\end{eqnarray}

Now, we use the induction hypothesis to each of the terms of the
right hand side of the above relation. This gives the identity we
write in the following long formula that should be understood in
this sense: the row labelled as (\ref{21}) is equal to the row
labelled as (\ref{29}), (\ref{22}) is equal to (\ref{30}) and so
on up to (\ref{28}) equal to (\ref{36}). In the next chain of
formulas $3l$ is a subindex for the $N$ first qubits of the third
group and $3r$ labels the last qubit (the $3n+3$-th) of this group
(now each group has $n+1$ qubits by the induction hypothesis).
Note that the forthcoming formula, although rather long, gives
already the desired answer straightforwardly. Thus, the above
relation equals to

\begin{eqnarray}
  \frac 12 \left[\frac{1}{2^n} \sum_{a_1,\cdots, a_n=0}^1\,|a_1\dots
a_n 0 a_{n+1}\dots
  a_{2n}0\rangle_{12}\left(\bigotimes_{k=a_{n+1}}^{a_{2n}}X^k\right)\,
  \left(\bigotimes_{l=a_{1}}^{a_{n}}Z^l\right) \sum_{i=0}^{2^n-1}
\alpha_{2i}\,|i_n\rangle_{3l}\,|0\rangle_{3r}\right.\nonumber\\\label{29}\end{eqnarray}
\begin{eqnarray}
   +\frac{1}{2^n} \sum_{a_1,\cdots, a_n=0}^1\,|a_1\dots a_n 0
a_{n+1}\dots
  a_{2n}1\rangle_{12}\left(\bigotimes_{k=a_{n+1}}^{a_{2n}}X^k\right)\,
  \left(\bigotimes_{l=a_{1}}^{a_{n}}Z^l\right) \sum_{i=0}^{2^n-1}
  \alpha_{2i+1}\,|i_n\rangle_{3l}\,|0\rangle_{3r}\nonumber\\
  \label{30}\end{eqnarray} \begin{eqnarray}
     +\frac{1}{2^n} \sum_{a_1,\cdots, a_n=0}^1\,|a_1\dots a_n 1
a_{n+1}\dots
  a_{2n}0\rangle_{12}\left(\bigotimes_{k=a_{n+1}}^{a_{2n}}X^k\right)\,
  \left(\bigotimes_{l=a_{1}}^{a_{n}}Z^l\right) \sum_{i=0}^{2^n-1}
  \alpha_{2i}\,|i_n\rangle_{3l}\,|0\rangle_{3r}\nonumber\\
  \label{31}\end{eqnarray} \begin{eqnarray}
     +\frac{1}{2^n} \sum_{a_1,\cdots, a_n=0}^1\,|a_1\dots a_n 1
a_{n+1}\dots
  a_{2n}1\rangle_{12}\left(\bigotimes_{k=a_{n+1}}^{a_{2n}}X^k\right)\,
  \left(\bigotimes_{l=a_{1}}^{a_{n}}Z^l\right) \sum_{i=0}^{2^n-1}
  \alpha_{2i+1}(-1)\,|i_n\rangle_{3l}\,|0\rangle_{3r}\nonumber\\
  \label{32}\end{eqnarray} \begin{eqnarray}
     +\frac{1}{2^n} \sum_{a_1,\cdots, a_n=0}^1\,|a_1\dots a_n 0
a_{n+1}\dots
  a_{2n}1\rangle_{12}\left(\bigotimes_{k=a_{n+1}}^{a_{2n}}X^k\right)\,
  \left(\bigotimes_{l=a_{1}}^{a_{n}}Z^l\right) \sum_{i=0}^{2^n-1}
  \alpha_{2i}\,|i_n\rangle_{3l}\,|1\rangle_{3r}\nonumber\\
  \label{33}\end{eqnarray} \begin{eqnarray}
     +\frac{1}{2^n} \sum_{a_1,\cdots, a_n=0}^1\,|a_1\dots a_n 0
a_{n+1}\dots
  a_{2n}0\rangle_{12}\left(\bigotimes_{k=a_{n+1}}^{a_{2n}}X^k\right)\,
  \left(\bigotimes_{l=a_{1}}^{a_{n}}Z^l\right) \sum_{i=0}^{2^n-1}
  \alpha_{2i+1}\,|i_n\rangle_{3l}\,|1\rangle_{3r}\nonumber\\
  \label{34}\end{eqnarray} \begin{eqnarray}
     +\frac{1}{2^n} \sum_{a_1,\cdots, a_n=0}^1\,|a_1\dots a_n 1
a_{n+1}\dots
  a_{2n}1\rangle_{12}\left(\bigotimes_{k=a_{n+1}}^{a_{2n}}X^k\right)\,
  \left(\bigotimes_{l=a_{1}}^{a_{n}}Z^l\right) \sum_{i=0}^{2^n-1}
  \alpha_{2i}\,|i_n\rangle_{3l}\,|1\rangle_{3r}\nonumber\\
  \label{35}\end{eqnarray} \begin{eqnarray}
    \left. +\frac{1}{2^n} \sum_{a_1,\cdots, a_n=0}^1\,|a_1\dots a_n 1
a_{n+1}\dots
  a_{2n}0\rangle_{12}\left(\bigotimes_{k=a_{n+1}}^{a_{2n}}X^k\right)\,
  \left(\bigotimes_{l=a_{1}}^{a_{n}}Z^l\right) \sum_{i=0}^{2^n-1}
\alpha_{2i+1}(-1)\,|i_n\rangle_{3l}\,|1\rangle_{3r}\right]\,.\nonumber\\
  \label{36}
\end{eqnarray}

Let us analyze the above sum. It contains four kinds of terms:

\begin{itemize}
    \item Two terms with $|0,0\rangle= |a_1\dots a_n 0 a_{n+1}\dots
  a_{2n}0\rangle$. These terms are (\ref{29}) and (\ref{34}). In
  fact, if we add up (\ref{29}) and (\ref{34}) we obtain a term of
  the form $2^{-n}\sum_{a_1,\dots,a_n=0}^1|a_1\dots a_n 0a_{n+1}\dots
a_{2n}0\rangle_{12}
  (\otimes_{k=a_{n+1}}^{a_{2n}}X^{k})(\otimes_{l=a_{1}}^{a_{n}}Z^{l})$
  times the following sum

  \begin{equation}\label{38}
\sum_{i=0}^{2^n-1}\alpha_{2i+1}|i_n\rangle_{3l}|1\rangle_{3r}+
\sum_{i=0}^{2^n-1}\alpha_{2i}|i_n\rangle_{3l}|0\rangle_{3r}
=\sum_{i=0}^{2^{n+1}-1}\alpha_i|i_{n+1}\rangle_3=|\psi_{n+1}\rangle\,.
\end{equation}
In this case, Bob's state is obviously
$(\otimes_{k=a_{n+1}}^{a_{2n}}X^{k})(\otimes_{l=a_{1}}^{a_{n}}Z^{l})|\psi_{n+1}\rangle$.

    \item Two terms with $|0,1\rangle= |a_1\dots a_n 0 a_{n+1}\dots
  a_{2n}1\rangle$, which are (\ref{30}) and (\ref{33}). In this
  case, the sum (\ref{38}) is changed into

  \begin{equation}\label{39}
\sum_{i=0}^{2^n-1}\alpha_{2i}|i_n\rangle_{3l}|1\rangle_{3r}+
\sum_{i=0}^{2^n-1}\alpha_{2i+1}|i_n\rangle_{3l}|0\rangle_{3r}
=\sum_{i=0}^{2^{n+1}-1}\alpha_i
X|i_{n+1}\rangle_3=X|\psi_{n+1}\rangle\,.
\end{equation}
Note that $X$ applies to the last qubit only and the other $n$
remain unchanged. Therefore, we should have rigorously written
$I^{\otimes n}\otimes X$ to denote the tensor product $n$ times
the identity operator and one time $X$, but we have written $X$
for simplicity.
    \item Two terms with $|1,0\rangle= |a_1\dots a_n 1 a_{n+1}\dots
  a_{2n}0\rangle$, which are (\ref{31}) and (\ref{36}). The sum
  gives here

  \begin{equation}\label{40}
    Z|\psi_{n+1}\rangle\,.
\end{equation}
As in the previous case, we have written $Z$ instead of
$I^{\otimes n}\otimes Z$ for simplicity.
    \item Two terms with $|1,1\rangle= |a_1\dots a_n 1 a_{n+1}\dots
  a_{2n}1\rangle$, which are (\ref{32}) and (\ref{35}). The sum is
  in this case

  \begin{equation}\label{41}
    XZ|\psi_{n+1}\rangle\,.
\end{equation}
Here, $XZ$ replaces $(I^{\otimes n}\otimes X)(I^{\otimes n}\otimes
Z)$.
\end{itemize}
Note the similarities with the $N=1$ case. Finally, we conclude
that the last long formula (\ref{29}-\ref{36}) is

\begin{equation}\label{42}
    \frac{1}{2^{n+1}}\sum_{a_1\cdots a_{n+1}=0}^1 |a_1\cdots
a_{2n+2}\rangle_{12}\left(\bigotimes_{k=a_{n+2}}^{a_{2n+2}}X^k\right)
\left(\bigotimes_{l=a_{1}}^{a_{n+1}}Z^l\right)|\psi_{n+1}\rangle_3\,.
\end{equation}
Thus, we have obtained equation (\ref{12}).

\subsection{Description of the teleportation
circuit.}\label{sec:DescripCircuit}

The above mathematical presentation can be summarize in terms of
what is call a {\it quantum circuit}. The circuit permits us a
visualization of the computational process.

The input in the teleportation circuit (see Figure 1), is given by
the $N$ qubit state $|\psi_N\rangle$ and the $2N$ ancillary qubits
all in the state $|0\rangle$. We readily see that

i.) The first set of $N$ ancillary qubits goes through Hadamard
gates, which are here represented by the symbol $\boxed{H}$ .

ii.) CNOT operations involve two qubits and are represented by a
dot $\bullet$ in the control qubit (the qubit that determines the
operation to be performed in the other qubit or target qubit), a
wire that connects the dot with the symbol $\oplus$ over the
target qubit.

iii.) Up to this point, all the operations are performed over the
ancillary qubits. A dashed line separates these operations and all
subsequent ones.

iv.) After the dashed line, we make CNOT operations on the $2N$
first qubits as shown in the circuit.

v.) The next operation is passing the first $N$ quibits thorugh
respective Hadamard gates.

vi.) A measurement process is carried out in the first $2N$
qubits. This is shown in the circuit by means of the symbol
corresponding to a measurement apparatus.

vii.) The horizontal dashed line separates between the Alice's
qubits (above of the line) from the Bob's qubits (below of the
line).

viii.) Finally, the box in the lower right corner indicates the
operation that Bob should perform in order to obtain the original
$N$ qubit state.

\begin{acknowledgments}
We want to express our gratitude to Mr. Guido Macchi and Dr.
Alfredo Caro for constant encouragement. We acknowledge partial
financial support to Junta de Castilla y Le\'on through Project
VA013C05 and the Ministry of Education of Spain (PR 2004-0080).
\end{acknowledgments}

\newpage
\section*{Figures}
\begin{eqnarray*}
\Qcircuit @C=.5em @R=.7em { & \qw & \qw & \qw & \qw & \push{|} \qw
& \ctrl{4} & \qw & \qw &
\gate{H} & \meter & \controlo \cw \cwx[1] \\
& \qw & \qw & \qw & \qw & \push{|} \qw & \qw & \ctrl{4} & \qw &
\gate{H} & \meter & \controlo \cw \cwx \\
\lstick{\ket{\psi_N}} & \push{\vdots} & \push{\rule{0em}{1em}} &
\push{\rule{0em}{1em}} & \push{\rule{0em}{1em}} & \push{|} &
\push{\rule{0em}{1em}} & \push{\rule{0em}{1em}} &
\push{\rule{0em}{1em}} &
\push{\vdots} & \push{\rule{0em}{1em}} & \cwx \\
& \qw & \qw & \qw & \qw & \push{|} \qw & \qw & \qw & \ctrl{4} &
\gate{H} & \meter & \controlo \cw \cwx \\
\lstick{\ket{0}} & \gate{H} & \ctrl{5} & \qw & \qw & \push{|} \qw
&
\targ & \qw & \qw & \qw & \meter & \controlo \cw \cwx \\
\lstick{\ket{0}} & \gate{H} & \qw & \ctrl{5} & \qw & \push{|} \qw
& \qw
& \targ & \qw & \qw & \meter & \controlo \cw \cwx \\
& \push{\vdots} & \push{\rule{0em}{1em}} & \push{\rule{0em}{1em}}
& \push{\rule{0em}{1em}} & \push{|} & \push{\rule{0em}{1em}} &
\push{\rule{0em}{1em}} & \push{\rule{0em}{1em}} & \push{\vdots} &
\push{\rule{0em}{1em}} & \cwx \\
\lstick{\ket{0}} & \gate{H} & \qw & \qw & \ctrl{5} & \push{|} \qw
& \qw & \qw & \targ & \qw
& \meter & \controlo \cw \cwx \cwx \\
& \push{\rule{0em}{1em}} & \push{\rule{0em}{1em}} &
\push{\rule{0em}{1em}} & \push{\rule{0em}{1em}} & \push{|} &
\push{-} & \push{-} & \push{-} & \push{--} & \push{--} &
\push{--\;--\;--\;--\;--\;--} \cwx &
\rstick{\mbox{Alice} \atop \mbox{Bob}} \\
\lstick{\ket{0}} & \qw & \targ & \qw & \qw & \push{|} \qw & \qw &
\qw & \qw & \qw & \qw &
\multigate{3}{\left(\!\!\bigotimes\limits_{k=M_{1}}^{M_{n}}{\!\!\!Z^k\!\!}
\right)\!\!\left(\!\!\bigotimes\limits_{k=M_{n+1}}^{M_{2n}}{\!\!\!X^k\!\!}\right)}
\cwx & \qw & \\
\lstick{\ket{0}} & \qw & \qw & \targ & \qw & \push{|} \qw & \qw &
\qw & \qw & \qw & \qw &
\ghost{\left(\!\!\bigotimes\limits_{k=M_{1}}^{M_{n}}{\!\!\!Z^k\!\!}\right)\!\!
\left(\!\!\bigotimes\limits_{k=M_{n+1}}^{M_{2n}}{\!\!\!X^k\!\!}\right)}
& \qw & \\
& \push{\vdots} & \push{\rule{0em}{1em}} & \push{\rule{0em}{1em}}
& \push{\rule{0em}{1em}} & \push{|} & \push{\rule{0em}{1em}} &
\push{\rule{0em}{1em}} & \push{\rule{0em}{1em}} & \push{\vdots} &
\push{\rule{0em}{1em}} & \push{\rule{0em}{1em}} &
\push{\rule{0em}{1em}} &
\rstick{\ket{\psi_N}} \\
\lstick{\ket{0}} & \qw & \qw & \qw & \targ & \push{|} \qw & \qw &
\qw & \qw & \qw & \qw &
\ghost{\left(\!\!\bigotimes\limits_{k=M_{1}}^{M_{N}}{\!\!\!Z^k\!\!}\right)
\!\!\left(\!\!\bigotimes\limits_{k=M_{N+1}}^{M_{2N}}{\!\!\!X^k\!\!}\right)}
& \qw & \\
& & & & & & & & & &\\
& & & & & & & & & &\\
& & & & & & & & & &\\
& & & & & & & & & &
 \mbox{\begin{small}FIG. 1: N qubit state
teleportation circuit\end{small}} }
\end{eqnarray*}


\begin{thebibliography}{99}
\bibitem{7} In fact the vectors
$|0\rangle$ and $|1\rangle$ represent the basis for any two level
system. Among them the spin $1/2$ is perhaps the most used. This
is why, we use the language of spin up and spin down.
\bibitem{8} We recall that the Pauli matrices are: $$
\sigma_x=\left(\begin{array}{cc}
  0 & 1 \\
  1 & 0 \\
\end{array}
\right)\hskip0.6cm;\hskip0.6cm \sigma_y=\left(
\begin{array}{cc}
  0 & i \\
  -i & 0 \\
\end{array}
\right)\hskip0.6cm ;\hskip0.6cm \sigma_z=\left(
\begin{array}{cc}
  1 & 0 \\
  0 & -1 \\
\end{array}
\right)\;.$$
\bibitem{9} In fact, any vector in ${\mathbb
C}^2$ has the matrix form given by $$ \left(\begin{array}{c}
  \alpha \\
  \beta \\
\end{array}\right)=\alpha\left(\begin{array}{c}
  1 \\
  0 \\
\end{array}\right)+\beta \left(\begin{array}{c}
  0 \\
  1 \\
\end{array}\right)\,,
$$
where $\alpha$ and $\beta$ are complex numbers. If we write as
usual $$ |0\rangle:= \left(
\begin{array}{c}
  1 \\
  0 \\
\end{array}
\right) \hskip0.6cm {\rm and} \hskip0.6cm |1\rangle:= \left(
\begin{array}{c}
  0 \\
  1 \\
\end{array}
\right)\,,
$$
we identify in a natural fashion the qubit
$\alpha|0\rangle+\beta|1\rangle$ with the element of ${\mathbb
C}^2$ $$ \left(
\begin{array}{c}
  \alpha \\
  \beta\\
\end{array}
\right)\,.
$$
If we have a system of $N$ qubits, the Hilbert space that we need
for such description is ${\mathbb C}^{2^N}$. Each vector of
${\mathbb C}^{2^N}$ (denoted as $|\psi_N\rangle$ or $\psi_N$) is a
column matrix with $2^N$ complex entries. Each basis vector can be
written in the form $|m_1 m_2\dots m_N\rangle$, where $m_i$,
$i=1,2,\dots,N$, are either $0$ or $1$ and is identified with a
column matrix with all components equal to zero except one which
gives $1$.
\bibitem{10} Note that
the CNOT operation is the quantum analog to the classical XOR
operation.
\bibitem{11} The
tensor product of two $2\times 2$ matrices is defined as follows:
$$
\left(
\begin{array}{cc}
  a & b \\
  c & d \\
\end{array}
\right)\otimes \left(
\begin{array}{cc}
  \alpha & \beta \\
  \gamma & \delta \\
\end{array}
\right)= \left(
\begin{array}{cc}
  a \, \left(
\begin{array}{cc}
  \alpha & \beta \\
  \gamma & \delta \\
\end{array}
\right)& b\, \left(
\begin{array}{cc}
  \alpha & \beta \\
  \gamma & \delta \\
\end{array}
\right) \\
  c\, \left(
\begin{array}{cc}
  \alpha & \beta \\
  \gamma & \delta \\
\end{array}
\right) & d\, \left(
\begin{array}{cc}
  \alpha & \beta \\
  \gamma & \delta \\
\end{array}
\right) \\
\end{array}
\right)= \left(
\begin{array}{cccc}
  a\alpha & a\beta & b\alpha & b\beta \\
  a\gamma & a\delta & b\gamma & b\delta \\
  c\alpha & c\beta & d\alpha & d\beta \\
  c\gamma & c\delta & d\gamma & d\delta \\
\end{array}
\right)\,.
$$
The generalization to tensor products of $m\times m$ matrices is
straightforward.
\end{thebibliography}
\end{document}